\documentstyle[prb,multicol,aps,epsfig]{revtex}
\begin{document}

\title{Critical current in Nb-Cu-Nb junctions with non-ideal interfaces}
\author{Y. Blum$^1$, A. Tsukernik$^2$, M. Karpovski$^1$, A. Palevski$^1$}
\address{
$^1$ School of Physics and Astronomy, Tel Aviv University, Tel Aviv
69978, Israel\\
$^2$ University Research Institute for Nanoscience and Nanotechnology,
Tel Aviv University, Tel Aviv 69978, Israel
}
\maketitle
\begin{abstract}
We report on experimental studies of superconductor (Nb) -normal metal
(Cu) -superconductor (Nb)
junctions with dirty interfaces between the different materials. By
using a set of simultaneously prepared samples, we
investigated the thickness dependence as well as the temperature
dependence of the critical currents in the junctions. Good agreement 
between the decay of the measured critical currents and theoretical
calculations was obtained without any fitting parameters.
\end{abstract}
\draft
\begin{multicols}{2}

\section{Introduction}

The non-dissipative current, as well as other superconducting
properties, are induced in the normal metal by an electron-hole
pair that is formed during the Andreev reflection at the
superconductor (S) -normal metal (N) interface \cite{Andreev}. The finite
length over which the electron-hole pair remains correlated determines
the characteristic decay length, $\xi_N$, of the superconducting
current in SNS junctions.
The normal-metal thickness, $L$, and temperature dependence of the super-current in the long
SNS junctions, $L>\xi_N$, has been thoroughly investigated over the
last few decades, both theoretically
\cite{Zaikin,Kupriyanov82,Kupriyanov88} and experimentally
\cite{Clarke,Courtois,Dubos}.

However, only in the case of junctions with clean interfaces, was reasonably good
agreement between the theory \cite{Zaikin} and the experiment
\cite{Dubos} found. In practice, there is no way to produce
vanishing boundary resistance. Nevertheless, the ratio of the latter to the
resistance of the normal metal in the junction could be made much
smaller than unity, as was the case for the lateral junctions in Ref. 7. In this limit, the theories developed for the ideal
interface junction are valid. 
It should be emphasized, that the high quality interface obtained during
{\it in situ} sequential deposition of two materials could result in
large values of the ratio provided that the resistance of the
normal-metal in the junction attains a low value. 
For non-ideal junctions, namely, for devices with vertical junctions,
the ratio of interface to normal-metal resistances is usually larger
than unity. Therefore, for such devices the theory of the ideal interface
limit \cite{Zaikin} is not applicable. The influence of boundary
scattering on the critical current in SNS junctions has been
considered theoretically in the past \cite{Kupriyanov82,Kupriyanov88},
However, so far, there is no
experimental data which quantitatively verifies
the theoretical prediction. 
The most remarkable difference between the theoretical predictions for
the ideal and non-ideal junctions is
reflected in the dependence of the critical superconducting current,
$I_C$, on the total junction resistance in the normal state, $R_j$. For the ideal
junctions, where normal-metal resistance is the main contributor to the
total $R_j$, $I_C \propto R_j^{-1}$. In the opposite limit, where the
main contribution to $R_j$ arises from the interface resistance,
the supper-current is proportional to $R_j^{-2}$.

This paper is devoted to experimental studies of the critical current
in vertical Nb-Cu-Nb long junctions with a large ratio of interface to
normal-metal resistances. We
demonstrate good quantitative agreement with the theory of Kupriyanov
{\it et al.} \cite{Kupriyanov88} in the appropriate limit.

\section{Theoretical background}
For the long SNS junction, the one dimensional Usadel equations
\cite{Usadel} are used with the appropriate boundary conditions. In the zero
approximation in the Matsubara frequencies, the critical current in a
junction with ideal interfaces can be expressed as follows \cite{Zaikin}:

\begin{equation}
I_c=\frac{64\pi k_B T}{eR_j}\frac{\Delta^2(T)}{[\pi k_B
  T+\Omega+(2\Omega(\pi k_B
  T+\Omega))^{1/2}]^2}\frac{L}{\xi_N}e^{-\frac{L}{\xi_N}}
\label{ideal}
\end{equation}

where $\Delta(T)$ is the
superconductor energy gap, $\Omega=\sqrt{(\pi k_B T)^2+\Delta^2(T))}$,
and
$\xi_N=\sqrt{\frac{\hbar D}{2\pi k_B T}}$ the thermal length.

In the case of non-ideal interfaces (transmission less than unity)
between the normal metal and superconductors, the Eilenberger
equations \cite{Elinberger} must first be solved for the vicinity of
the boundary, after which it is possible to write the boundary conditions for the Usadel equations.

Following Ref. 4, the resistivity of the structure in the normal state can be expressed as 

\begin{equation}
R_j=\frac{L}{\sigma_NA}(1+2\Gamma_B)
\label{Gamma}
\end{equation}

where $\sigma_N$ is the normal metal conductivity, and
$\Gamma_B=\frac{\xi_N^*\gamma_B}{L}$. Here, $\gamma_B$ is a parameter
related to the transparency of the interface.

In the limit of SN boundary with small transparency
$\gamma_B>>(\frac{T_c}{T})^{1/2}$,
the critical current in the first order of the Matsubara frequency is given by \cite{Kupriyanov88}:

\begin{equation}
I_c=\frac{4\pi k_BT_c}{eR_j}\frac{1+2\Gamma_B}{\gamma_B^2}
\frac{\Delta^2(T)}{(\pi
  k_BT)^2+\Delta^2(T)}\frac{L}{\xi_N}e^{-\frac{L}{\xi_N}}
\label{Kupriyanov}
\end{equation}

where $T_c$ is the superconductor critical temperature. For the low transparency junctions, namely $\Gamma_B>>1$, the critical
current can be expressed in terms of measurable quantities as follows:

\begin{equation}
I_c=\frac{16\pi k_BT_c}{eA_j}\frac{\rho_N}{R_j^2}
\frac{\Delta^2(T)}{(\pi k_BT)^2+\Delta^2(T)}\frac{{\xi_N^*}^2}{\xi_N}
e^{-\frac{L}{\xi_N}}
\label{Yuval}
\end{equation}

where $\rho_N$ is the normal metal resistivity and
$\xi_N^*=\sqrt{\frac{\hbar D}{2\pi k_B T_c}}$. Eq. \ref{Yuval} is
valid only for $L\gg \xi_N$. For $L\geq \xi_N$ the contribution of higher
harmonics of the Matsubara frequencies should be included (see
Eq. (34) of Ref. 4).

\section{Sample preparation and measurement set-up}

\begin{figure}
      \epsfig{file=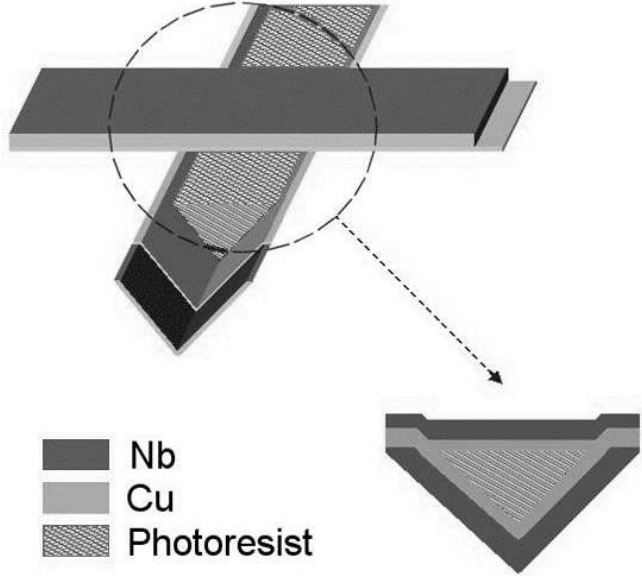, angle=0,  width=\linewidth}
      \caption{Schematic drawing of a sample. A cross section of 
       the structure is given on the right side.\label{fig:Sample.eps}}
\end{figure}

The samples were fabricated as long triangular prisms. Two SNS junctions
were 
created along the prism in a SQUID like geometry as shown
schematically in Fig.\ref{fig:Sample.eps}. This
design allows the observation of Little-Parks 
oscillations \cite{LP}  which give two advantages. First, it is possible to
ensure zero magnetic flux through the junctions. Second,
accurate measurements of the normal resistance and the cross section
area of the junctions is possible. 

The SNS junctions were fabricated on top of V-shaped
substrates, produced by wet-chemical crystallographic etching
\cite{Tsukernik} of GaAs planar substrates.

The V-groove substrate was covered with Nb and Cu films, which produced
the two bottom side-walls of the prism. Nb films were sputtered using
a magnetron gun and covered {\it in situ} 
with a Cu layer by thermal evaporation, to prevent Nb
oxidation. An additional Cu layer was e-gun evaporated in a separate vacuum chamber.
These layers were completely
covered by photoresist. Oxygen plasma etching was applied in order to
expose the top parts of the V-groove. The subsequent depositions of
the Cu
and Nb films, as described above, completed the fabrication of the prism samples. 
The crossing geometry of the top and the bottom films allowed four
terminal measurement of the junction below the critical temperature 
of the Nb films.

Nine sets of samples were prepared simultaneously in order to assure
identical interfaces at all 
junctions. The samples of each set had an identically thick Cu
layer.
Variation of the Cu layer
thickness between the different sets was obtained by a specially designed shutter, which
exposed the samples in sequence, so that every set was
exposed to the evaporating Cu for additional fragments of time.
Thus, the only difference between the sets was the
thickness of the Cu layer. 
We prepared separate Cu films
simultaneously with the deposition of the Cu in the junctions. 
The measured resistivity of these Cu films at 4.2K was $\rho_N=5.5 \cdot
10^{-9} \Omega m$.

The thickness of each Nb layer was
2000\AA, whereas the Cu thickness varied from $L=2000\AA~$ to $L=10000\AA$. 
The area of the junction, $A_j$, relevant for the critical current is determined by
the width of the Nb strips, $W_{Nb}=10 \mu m$, and the width of the exposed top part of
the V-groove, $W_{top}$. The cross-section of the junction
relevant for the flux dependence of the critical current in the
junction is given by $LW_{top}$. The overall flux dependence of the
critical current in the
SQUID sample is determined both by the area of the junction and the
total cross section of the SQUID ($A_{SQUID}=6.5 \mu m^2$), as given
in the following expression \cite{Duzer}:
\begin{equation}
I_{c_{total}}=2I_{c_j}\left|{\cos \frac{\pi \Phi}{\Phi_0}\left[
\frac{\sin(\pi \Phi_j/\Phi_0)}{\pi \Phi_j/\Phi_0}\right]}\right|
\label{Duzer}
\end{equation}
where $I_{c_j}$ is the critical current of one junction, while $\Phi$
and $\Phi_j$ are the flux in the SQUID and junction respectively.
  
\begin{figure}
  \begin{center}
      \epsfig{file=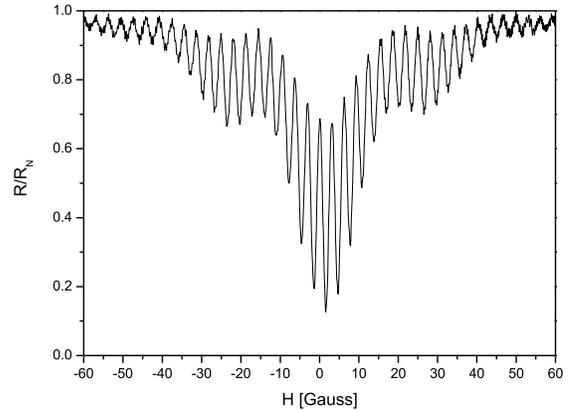, angle=0, width=\linewidth}
      \caption{Typical magnetoresistance plot as measured in a
         sample  with a layer of 8000\AA Cu.\label{fig:Osci.eps}}
  \end{center}
\end{figure}

The measurements were performed in a $^4$He cryostat in the range
1.3K-4.2K. The critical current was measured by passing a DC current with a
small AC modulation through the sample. The AC voltage, which appeared
above the critical DC current, was picked up by a lock-in
amplifier operated in transformer mode.

\begin{figure}
  \begin{center}
      \epsfig{file=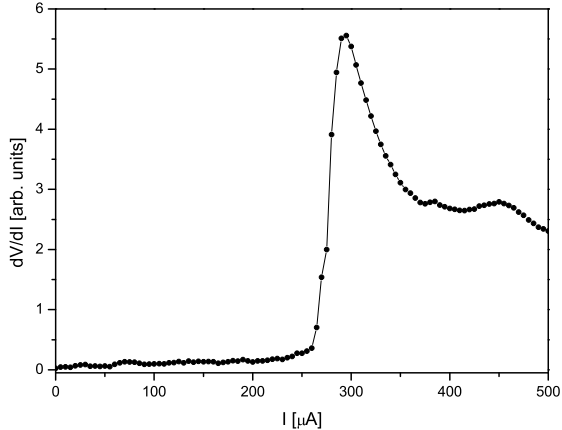, angle=0, width=\linewidth}
      \caption{Typical variation of the differential resistance of a
        SNS junction. The onset of the peak was defined as the
        critical current of the junction.\label{fig:dVdI.eps}}
  \end{center}
\end{figure}

\section{Experimental data and analysis}
We start the description of our data with the magnetoresistance
oscillations of the samples. Fig.\ref{fig:Osci.eps} shows a typical
magnetoresistance plot for a sample with a Cu thickness of
$L=8000\AA$. 
The short oscillation period, $\Delta H=3.1$ Gauss, corresponds to the
area $6.5 \mu m^2$ enclosed by the
SQUID, which is consistent with our designed geometry of the samples. 
The beating modulation with larger period and attenuation of the oscillations
arises from additional Fraunhofer oscillations of the critical current
in each junction (see Eq. \ref{Duzer}), with typical area $LW_{top}$.
Therefore, the analysis of the magnetoresistance oscillations allows
us to extract $W_{top}=1.2 \mu m$, resulting in $A_j=12 \mu m^2$. Thus,
magnetoresistance oscillations
determine the geometrical dimensions of our samples, given
in the previous section in the parenthesis.

The critical currents in our samples were measured at temperatures in
the range of 1.3K-4.2K using the setup described above.
A typical differential resistance of a sample vs. DC current is
plotted in Fig.\ref{fig:dVdI.eps}, which clearly indicates the sharp
onset of dissipation at $I=I_c$. 

Some of the parameters which appear in Eq. \ref{ideal} and Eq. \ref{Yuval},
namely $R_j=19m \Omega, T_c=8.0K$ and $\rho_{Cu}=5.5 \cdot 10^{-9}
\Omega m$, are directly
measured in our experiment.
The rest of the parameters, namely $\xi_N$, $\xi_N^*$
and $\Delta$ can be calculated.
The diffusion coefficient, which appears in $\xi_N$ and $\xi_N^*$,
can be calculated from the measured resistivity of
the Cu, using the relation $D=(e^2N(\epsilon_F)\rho)^{-1}$, where
$N(\epsilon_F)=1.56 \cdot 10^{47} J^{-1}$ is the density of states at
the Fermi energy for Cu \cite{Pierre}. 
From the latter, we obtain $D=4.5 \cdot
10^{-2} m^2/sec$.
The energy gap is calculated
using the known relation for Nb\cite{Duzer} $\Delta=1.9k_BT_c=1.3 meV$.
Thus, there are no free parameters entering the theoretical
expressions.

We should keep in mind that Eq. \ref{Kupriyanov} is only valid for
large values of $\gamma_B$.
The value of $\gamma_B$ calculated from Eq.\ref{Gamma} is about 500, and consequently,
$\Gamma_B$ varies between $\sim 50$ for $L=9000\AA$ to $\sim 200$ for
$L=2000\AA$. These values justify the applicability of both
Eq.\ref{Kupriyanov} and Eq.\ref{Yuval}.

\begin{figure}
  \begin{center}
      \epsfig{file=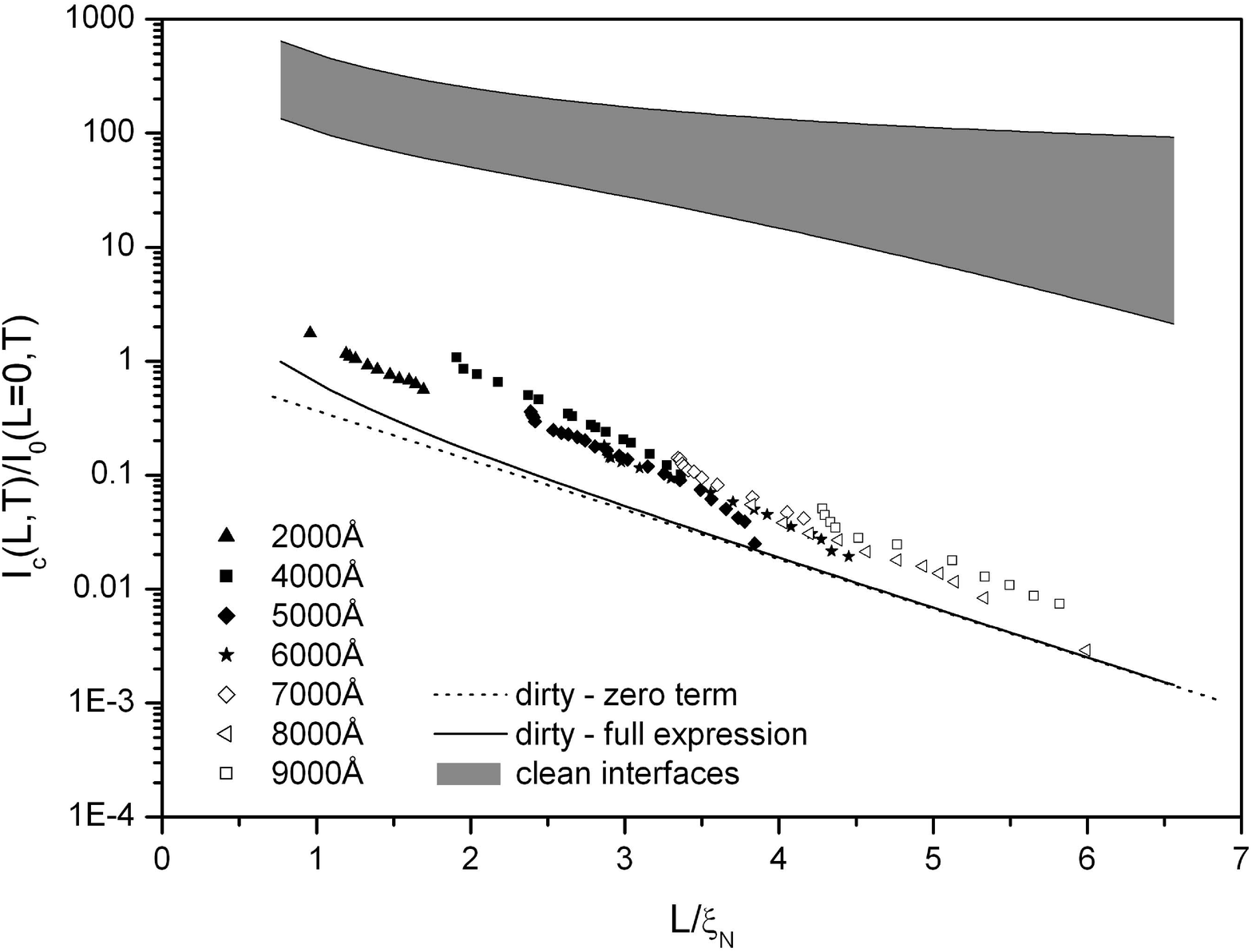, angle=0, width=\linewidth, height=8cm}
      \caption{Universal curve of the critical currents (dimensionless
        units), and the theoretical fits with the parameters described
        above. The dashed line represents Eq. \protect\ref{Yuval}, the
        solid line represents the infinite series (Eq. 34 in
        Ref. 4) and the gray area represents
        the theory for clean interface junctions 
        \protect\cite{Zaikin} with thickness
        between $2000\AA$ and $9000\AA$.\label{fig:Universal.eps}}
  \end{center}
\end{figure}

Since our aim is to compare our experimental data with the theory of
Kupriyanov {\it et. al}\cite{Kupriyanov88}, we choose a dimensionless frame of axis
($L/\xi_N$, $I_c(T,L)/I_0(T)$), where  $I_0(T)=\frac{16\pi k_BT_c}{eA}\frac{\rho_N}{R_j^2}
\frac{\Delta^2(T)}{(\pi k_BT)^2+\Delta^2(T)}\frac{{\xi_N^*}^2}{\xi_N}$. 
These axes are natural for the $I_c$ variation
since the leading term in the Matsubara frequency expansion
(Eq. \ref{Yuval}) is just a simple exponential dependence, and
therefore all our data points, presented in this frame of axis, should
collapse onto a single exponential curve. Fig. \ref{fig:Universal.eps} shows all
our data points which quite closely follow (within a factor of two) the
theory of Kupriyanof {\it et. al}. The dashed line in
Fig. \ref{fig:Universal.eps} corresponds to Eq. \ref{Yuval}, while the solid
line accounts for all Matsubara frequencies (Eq. 43 in Ref. 4).

The consistency with the latter
theory is even more striking when the data is compared with the theory
developed for a junction with clean interfaces \cite{Zaikin}. 
Since for this theory, the chosen frame of axis is not natural, our
data should fall into the gray area. The upper border of the area
corresponds to $L=2000\AA$ while the lower border corresponds to $L=9000\AA$. 
Our data deviates by
about three orders of magnitude from the theory of Zaikin {\it et al.}

It should be noted that there are no free parameters in the
evaluation of the theoretical formula. The origin of the deviation 
of our data from the theoretical curve is not clear, and could arise
from several sources. 
The estimate of the area from the Frounhofer oscillations has at least
$\pm20\%$ error. 
The smaller area would increase $I_0$ and therefore, improve the fit to
the theory. 
The additional source of the error could arise from the the value of $\rho$ in the Cu layer.
As explained above, it was determined
from the resistivity measurement of the Cu film in the planar geometry.
Although the small grain Cu films usually do not show significant
anisotropy, 
one should bear in mind that the deposition of our Cu films was made
with two 
interaptions (when moving the samples from the sputtering chamber to
the e-gun deposition
 chamber and back). These additional interfaces with partially
 oxidized Cu could 
make the resistivity of the Cu very anisotropic, being much larger in
the direction 
perpendicular to the interface than the measured values in planar
geometry. 
We have no experimental way of verifying this.
Apart from increasing $I_0$, and therefore improving the fit, the additional
interfaces could make some non-trivial changes to the boundary
conditions 
of the Usadel equations, which were not accounted for by the theory.

\section{Conclusions}
We demonstrated that the theory\cite{Kupriyanov88} of the critical
current in a SNS junction with non-ideal interfaces closely follows 
our experimental data. Keeping in mind that we have not used any
fitting parameters, we find the agreement quite amazing. 
We claim that for vertical junctions with a clean
normal-metal (even for those prepared {\it in situ}) the theory of Kupriyanov {\it et. al}\cite{Kupriyanov88} is
the relevant one. The theory developed for ideal-interface 
junctions\cite{Zaikin} may be applicable
either for lateral junctions or for vertical junctions containing a
strongly disordered metal, since only in these cases, the interface
resistance could become negligible relative to the resistance of
the normal metal.

\begin{acknowledgments}
We would like to thank A. Aharony, O. Entin-Wohlman, Y. Imry,
K. Kikoin, Z. Ovadyahu, A. Schiller, G. Sch\"on, K. Efetov and
A. Gerber for fruitful discussions. 
This research was supported by a Grant from G.I.F., the
German-Israeli Foundation for Scientific Research and Development as
well as the Israel Science Foundation founded by the Israel
Academy of Sciences and Humanities --- Center of
Excellence Program.
\end{acknowledgments}

\end{multicols}
\end{document}